\documentclass[11pt,twoside]{extarticle}

\usepackage{geometry}
\usepackage{fancyhdr}
\usepackage{amssymb}
\usepackage{mathrsfs}
\usepackage{mathtools}
\usepackage{centernot}
\usepackage{soul}
\usepackage{xcolor}

\usepackage[labelfont={bf}, textfont={small}, width=230pt]{caption}
\usepackage{subcaption}
\usepackage[export]{adjustbox}
\usepackage{tikz}

\usepackage[narrowiints]{kpfonts}
\DeclareFontFamily{U}{matha}{\hyphenchar\font45}
\DeclareFontShape{U}{matha}{m}{n}{
      <5> <6> <7> <8> <9> <10> gen * matha
      <10.95> matha10 <12> <14.4> <17.28> <20.74> <24.88> matha12
      }{}
\DeclareSymbolFont{matha}{U}{matha}{m}{n}
\DeclareFontSubstitution{U}{matha}{m}{n}
\DeclareMathSymbol{\lc}{4}{matha}{"74}
\DeclareMathSymbol{\rc}{5}{matha}{"75}

\usepackage[affil-it]{authblk}

\usepackage{abstract}

\usepackage{titlesec}
\titleformat*{\section}{\large\bfseries}
\numberwithin{equation}{section}

\usepackage[sorting=none, citestyle=numeric-comp, bibstyle=ieee]{biblatex}
\usepackage[colorlinks=true, linkcolor=blue, urlcolor=blue, citecolor=blue]{hyperref}

\newcommand{\articletitle}{
	Uncertainty and Wigner negativity in Hilbert-space classical mechanics
}

\newcommand{\shorttitle}{
	Uncertainty and Wigner negativity in Hilbert-space classical mechanics
}

\title{\Large{\articletitle}}

\author{
	Mustafa Amin
	\thanks{\href{mailto:m.amin@uleth.ca}{m.amin@uleth.ca}}
}

\affil{ \small
	Department of Physics \& Astronomy, University of Lethbridge, Lethbridge, Alberta, Canada
}

\date{}

\newcommand{\wt}{\widetilde}
\newcommand{\wh}{\widehat}
\newcommand{\wht}[1]{\widehat{\widetilde{#1}}}
\newcommand{\ve}{\varepsilon}
\newcommand{\Tr}{\text{Tr\,}}

\begin{document}

\maketitle

\begin{abstract}
	Classical mechanics, in the Koopman-von Neumann formulation, is described in Hilbert space.  It is shown here that classical canonical transformations are generated by Hermitian operators that are in general noncommutative.  This naturally brings about uncertainty relations inherent in classical mechanics, for example between position and the generator of space translations, between momentum and the generator of momentum translations, and between dynamical time and the Liouvillian, to name a few.  Further, it is shown that the Wigner representation produces a quasi-probability distribution that can take on negative values.  Thus, two of the hallmark features of quantum mechanics are reproduced, and become apparent, in a Hilbert-space formulation of classical mechanics.
\end{abstract}

\section{Introduction}\label{sec:intro}

Quantum mechanics and classical mechanics are different, but identifying the exact point at which the two theories diverge is illusive.  Looking at the usual representations of classical and quantum mechanics, one might be at a loss as to where the differences begin and where they end, and led to believe that quantum is wholly and qualitatively different from classical.  On the other hand, some similarities between the two show themselves enough to hint at a core that is common to both.  Prominent among these is the fundamental role played by the Poisson bracket in the classical case and the commutator in the quantum.  Both binary operations have almost identical algebraic properties, and both are central to the structure of their respective theory.  Is this more than just a curiosity?  Alas, a ``language'' barrier stands between us and a satisfying characterization of the common core, and the point of divergence.  A meaningful and robust comparison of concepts can take place only when unburdened and undistracted by difference in formulation.

It turns out that classical theory shares more with quantum theory than was traditionally believed.  If the state of reality, or ``ontic'' state, of a classical system is not fully known, then we can define our state of knowledge, or ``epistemic'' state, of the system as a probability distribution over the ontic states, and we resort to a statistical description of the system.  An example of an epistemic state in classical mechanics is the Liouville distribution on phase space, and the theory is statistical mechanics.  If the Liouville distribution is a delta function (representing a single point on phase-space), then it describes an ontic state.  It has been shown in Refs.~\cite{spekkens_defense_2007, bartlett_reconstruction_2012, spekkens_quasi-quantization_2016} that if we consider a statistical classical theory along with an epistemic restriction (a restriction on our knowledge), limiting the kinds of probability distributions available for us, then many concepts associated with quantum theory emerge.  Reference~\cite{spekkens_quasi-quantization_2016} lists noncommutativity, coherent superposition, collapse, complementarity, no-cloning, interference, teleportation, entanglement and others as features reproducible in a classical theory with an epistemic restriction.  In Ref.~\cite{kirkpatrick_quantal_2003} it was even argued that incompatibility of variables and associated phenomena arise in a deck of playing cards.

These findings, among others, demonstrate that a large part of what is usually believed to be characteristic of quantum mechanics is not so, and can be present in a classical theory.  A caveat in the examples above, though, is that they rely on a set of constructed rules.  In the case of the deck of cards in Ref.~\cite{kirkpatrick_quantal_2003}, it is obvious that certain game rules must be respected in order to reproduce the results; cheating is not allowed.  In the case of Refs.~\cite{bartlett_reconstruction_2012, spekkens_quasi-quantization_2016}, ordinary Liouville mechanics is used, but with an added extra-mechanical epistemic restriction.  In the words of Ref.~\cite{spekkens_quasi-quantization_2016}, ``Phenomena arising in an [epistemically-restricted] theory might still be considered to exhibit a type of nonclassicality insofar as an epistemic restriction is, strictly speaking, an assumption that goes beyond classical physics.''  Indeed, one may wonder, by what mechanism is knowledge restricted, if not another law of mechanics?

Here, we show that ordinary classical mechanics already possesses the structure to reproduce features traditionally thought to be quantum.  Namely, uncertainty and Wigner negativity.  The key to understanding this seemingly surprising, but in hindsight mundane, statement is to recognize that in addition to phase-space variables classical mechanics also contains generators of transformations as dynamical variables.

The starting point for this realization is the Koopman-von Neumann formulation of classical mechanics~\cite{Koopman1931, neumann_1932}.  There, a complex probability amplitude is defined such that its modulus-square is the Liouville probability distribution.\footnote{
	We prefer the term ``probability amplitude'' to ``wavefunction'' to emphasize its probabilistic role.  We will not be addressing wave-like phenomena.
}
In the Hilbert space of probability amplitudes, it is natural to consider operators that are in general noncommutative.  Operators that are Hermitian possess a complete set of eigenstates and are, in principle, observable.  Noncommutative dynamical variables will then give rise to many ``quantum-like'' phenomena.  These variables have been known to exist in the Koopman-von Neumann formulation (see, for example, Refs.~\cite{Mauro2002, Gozzi2004, Bondar2012, Wilczek2015}).  There has been an aura of mystery about them, and they were often dubbed ``extra'', ``auxiliary'', ``unphysical'', or ``unobservable.''  What do they represent, and what does it mean to have noncommutativity in a classical theory?  We show in Sec.~\ref{sec:cantran} that there is no mystery here.  The noncommutative variables present in the theory are generators of canonical transformations.  We attempt in Sec.~\ref{sec:postulates} to axiomatize the approach in order to establish a complete equivalence between the KvN formulation and ordinary classical mechanics in phase space, and so argue that the results obtained do, in fact, arise from classical mechanics without modification.

In quantum mechanics, position may be treated as fundamental and the generator of translation in position (up to a constant factor) is given the name ``momentum.''  Conversely, the position operator is the generator of translation in the aforementioned momentum.  By contrast, in Hilbert-space classical mechanics, both position and momentum are considered fundamental and they both have their separate generators of translation.  It is the existence of these pairs of variables and their associated generators of transformation that originates noncommutativity in both theories.

The unification of language used for classical and quantum mechanics allows us to compare the two without worrying about mismatching concepts.  For example, in both theories observables are represented by Hermitian operators and states by state operators.  Composite systems are described via tensor product of the Hilbert spaces of individual degrees of freedom.  Uncertainty relations and other features are derived in classical mechanics in the same way they are in quantum mechanics.  An example of a classical uncertainty relation between the dynamical time (see Ref.~\cite{bhamathi_time_2003}) and the Liouvillian has been discussed in Ref.~\cite{amin_solutions_2025}.

The key differences between classical and quantum mechanics, as discussed in Sec.~\ref{sec:postulates}, is that the former requires two Hilbert spaces per degree of freedom (one for position and one for momentum), and that only canonical transformations are allowed.  These are the conditions for the classical Hilbert space formulation to be equivalent to the ordinary phase-space formulation, where position and momentum are both treated as fundamental, and where the dynamics is preserved only under canonical transformation.  This, given the ``quantum-like'' results obtained, shows that some qualitative aspects of quantum mechanics can be derived from an underlying, decidedly ontic model: classical mechanics in phase space.

Could the aforementioned unified language have been that of phase space instead of Hilbert space?  In principle, yes, but that effort would have faced a major distraction: the notion that the quantum Wigner function is the analogue to the classical Liouville distribution.  We argue here that this is not the case.

Wigner brought quantum mechanics to phase space~\cite{wigner_quantum_1932}.  There, the state is represented as the Wigner quasi-probability distribution (also called Wigner function) and the observables are ordinary phase-space functions (for background see Refs.~\cite{Hancock2004, Zachos2005, Case2008, Curtright2014}).  This, it is hoped, facilitates comparison with classical mechanics where the state is represented as the Liouville probability distribution~\cite{Sudarshan2016, balescu_equilibrium_1975}.  Despite this unification of arena, concepts are still qualitatively different.  First, the Wigner distribution is not comparable to the Liouville distribution since the former is a quasi-probability that takes on negative values (Wigner negativity), while the latter is a true probability distribution.  True probability distributions in standard quantum mechanics are found as marginals of the Wigner function.  By contrast, a marginal of the Liouville distribution is a partial probability distribution, incapable of providing information regarding all variables.  Second, even though quantum observables are represented like their classical counterparts as phase-space functions, they are composed differently: classical observables use ordinary point-wise multiplication while quantum ones use the noncommutative star product, which is isomorphic to the Hilbert-space operator product~\cite{Hancock2004, Zachos2005, Curtright2014}.  It may be argued that classical mechanics should be obtained from quantum mechanics in this representation by taking the limit $\hbar \to 0$ where the star product becomes ordinary multiplication and the quantum equation of motion, the Moyal equation, produces the classical Liouville equation.  This, however, turned out not to be true~\cite{takahashi_distribution_1989, ballentine_quantum_2014}.  Further, the quantum phase-space formulation should be equivalent to its Hilbert-space formulation.  If we are interested in complete analogy between the mathematical language of the two theories, shouldn't classical mechanics too have a Hilbert-space formulation equivalent to its phase-space formulation?

In Hilbert space, it is easier to recognise the existence of a whole other class of dynamical variables, the generators of canonical transformations, and their noncommutativity.  Then, as shown in Sec.~\ref{sec:wigner}, a classical Wigner function is defined over a \textit{doubled} phase-space composed of position with its generator of translations, and momentum with its generator of translations.  It is shown that this classical Wigner function is a quasi-probability distribution, admitting negative values, like its quantum counterpart.  The Liouville probability distribution is obtained as a marginal of the classical Wigner function.  As such, a unification of language in phase space is possible, but not in the way usually expected.

This paper is organised as follows.  Section~\ref{sec:probamp} briefly introduces Liouville mechanics and its Koopman-von Neumann Hilbert-space counterpart.  Section~\ref{sec:cantran} details the implementation of canonical transformations on Hilbert space as unitary transformations generated by the so-called tilde-variables.  In Sec.~\ref{sec:tv} we discuss algebraic properties and representations of tilde-variables, as well as the fundamental commutation relations involving them and phase-space variables.  With the preliminaries out of the way, Sec.~\ref{sec:postulates} presents four postulates for classical mechanics in Hilbert space and contrasts them with their quantum analogues.  In Sec.~\ref{sec:uncertainty} we discuss the existence of classical uncertainty relations and their meaning in the ordinary phase-space formulation.  Finally, in Sec.~\ref{sec:wigner} we explore the Wigner representation of the Koopman-von Neumann formulation, discuss classical Wigner negativity, and relate to the quantum Wigner representation.  Section~\ref{sec:conc} concludes the paper.

\section{Probability amplitudes}\label{sec:probamp}

Classical systems are often described in terms of their positions and momenta which evolve over time according to Hamilton's equations.  It is an observational fact that the prediction (or retrodiction) of the motion of a classical particle needs both position and momentum in order to fully describe the mechanics of the system.  From here on, we will focus on systems of one degree of freedom $q$ and its associated momentum $p$ to simplify the presentation.

When precise values of $q$ and $p$ are not available, a probability distribution over them may be used for the prediction of the motion.  The equation of motion for the probability distribution $\rho_{_L}(q,p,t)$ is called the Liouville equation
\begin{align}\label{eq:liouvilleeq}
	\partial_t \rho_{_L} + {\lc}\rho_{_L}, H{\rc} = 0~.
\end{align}
Here, $\partial_x$ denotes partial derivative with respect to $x$, and $t$ is time.  The probability distribution $\rho_{_L}$ is called the Liouville distribution and, as a probability distribution, is real, nonnegative and normalized.  The Hamiltonian $H(q, p, t)$ is the physical quantity describing the dynamics of the system.  The Poisson bracket ${\lc}\cdot, \cdot{\rc}$ is discussed in Sec.~\ref{sec:cantran}.  The study of mechanics, emphasizing $\rho_{_L}$ and its equation of motion~\eqref{eq:liouvilleeq}, is sometimes called Liouville mechanics.  There, $\rho_{_L}$ evolves in time while $q$ and $p$ are held constant and form a space called phase space, where $\rho_{_L}$ moves.  This is analogous to the Schrödinger picture of quantum mechanics, as opposed to the Heisenberg picture where time resides in $q$ and $p$ which evolve according to Hamilton's equations (see~\cite{Sudarshan2016, balescu_equilibrium_1975}).  Equation~\eqref{eq:liouvilleeq} is central to statistical mechanics, which subsumes ordinary classical mechanics in the case where $\rho_{_L}$ is a delta function.  Liouville mechanics is an equivalent restatement of classical Hamiltonian mechanics.

In quantum mechanics (QM), the Schrödinger equation of motion describes not the evolution of a probability distribution, as the Liouville equation does for classical mechanics (CM).  Rather, the main quantity of interest in QM is a \textit{probability amplitude}.  That is, a quantity the modulus-square of which is a probability distribution.  Whereas the probability distribution is real, nonnegative and normalized, the probability amplitude is complex.  The space of all probability amplitudes constitutes the space of states for a quantum system; it is a (rigged) Hilbert space~\cite{ballentine_quantum_2014}.

A Hilbert-space formulation of classical mechanics, where the Liouville distribution is obtained as the modulus-square of a complex probability amplitude, was pioneered by Koopman in 1931, and von Neumann in 1932~\cite{Koopman1931, neumann_1932}.  The formulation, along with contributions from other scientists, is often referred to as the Koopman-von Neumann (KvN) formulation of classical mechanics.  For background, see Refs.~\cite{Mauro2002, Gozzi2004, Bondar2012, Wilczek2015, Klein2017, Bondar2019, amin_solutions_2025} and references therein.  The equation of motion for the probability amplitude $\chi(q,p,t)$, which we refer to as the KvN equation, is derived from that of the Liouville distribution $\rho_{_L} = |\chi|^2$ as
\begin{align}\label{eq:KvNequation}
	i\hbar \partial_t \chi = \wt{H} \chi~.
\end{align}
The ``tilde-Hamiltonian'' operator $\wt{H} \coloneqq i\hbar {\lc}H, \cdot{\rc} + \alpha_H(q,p,t)$ is a linear, Hermitian operator that is a generalization of the Liouvillian operator~\cite{amin_solutions_2025}, and $\alpha_H$ is an arbitrary function.  We will use ``tilde-Hamiltonian'' and ``Liouvillian'' interchangeably throughout.  Section~\ref{sec:cantran} contains a derivation of the tilde-Hamiltonian and similar operators as generators of classical canonical transformations.  Equation~\eqref{eq:KvNequation} is deliberately cast in a form similar to that of the Schrödinger equation to facilitate comparison.  At this point, the main differences between this classical equation and the quantum one is that the former concerns a function of $q$ and $p$, and that $\wt{H}$ takes form different from the usual quantum Hamiltonian.

Let us use a bra-ket notation for a more general treatment.  Let $|\chi \rangle$ denote a vector in the classical Hilbert space $\mathcal{H}$, and let $|q, p\rangle$ denote a vector corresponding to a state of definite $q$ and $p$.  The probability amplitude in the $(q,p)$-representation is defined as the inner product
\begin{align}\label{eq:qpchibraket}
	\chi(q,p,t) \coloneqq \langle q,p | \chi \rangle~.
\end{align}
We assume the completeness and (Dirac) orthonormality of the $|q,p\rangle$ basis
\begin{align}\label{eq:qpcompleteness}
	\int_{\underline{q}}^{\overline{q}} \int_{\underline{p}}^{\overline{p}} dq dp \,
	|q,p \rangle \langle q,p| = 1~, \quad \langle q',p' | q,p \rangle = \delta(q' - q) \delta(p' - p)~,
\end{align}
where $(\underline{q}, \overline{q})$ and $(\underline{p}, \overline{p})$ denote the lower and upper limits on $q$ and $p$.  In bra-ket notation, the KvN equation~\eqref{eq:KvNequation} is
\begin{align}\label{eq:KvNketequation}
	i\hbar \frac{d}{dt} |\chi\rangle = \wht{H} |\chi\rangle~,
\end{align}
with $\langle q,p| \frac{d}{dt} |\chi\rangle = \partial_t \chi(q,p,t)$ and $\langle q,p | \wht{H} | \chi \rangle = \wt{H} \chi(q,pt)$.

Further, let us define the basis vector $|q,p\rangle$ as the tensor product of a vector $|q\rangle$ in Hilbert space $\mathcal{H}_1$ and a vector $|p\rangle$ in Hilbert space $\mathcal{H}_2$
\begin{align}\label{eq:qptensor}
	|q,p\rangle \coloneqq |q\rangle \circledtimes |p\rangle~.
\end{align}
Similarly to $|q,p\rangle$, we assume the completeness and orthonormality of both bases in their respective spaces
\begin{align}\label{eq:qandpcompleteness}
	\int_{\underline{q}}^{\overline{q}} dq \, |q\rangle \langle q | &= 1~, \quad \langle q' | q \rangle = \delta(q' - q)~,\\
	\int_{\underline{p}}^{\overline{p}} dp \, |p\rangle \langle p | &= 1~, \quad \langle p' | p \rangle = \delta(p' - p)~.
\end{align}
Separate probability amplitudes $\chi^{(1)}(q,t)$ and $\chi^{(2)}(p,t)$ are defined as the inner products
\begin{align}\label{eq:qchipchibrakets}
	\chi^{(1)}(q,t) \coloneqq \langle q | \chi^{(1)} \rangle~, \quad
	\chi^{(2)}(p,t) \coloneqq \langle p | \chi^{(2)} \rangle~.
\end{align}
The existence of spaces $\mathcal{H}_1$ and $\mathcal{H}_2$ is a reflection of the fact that it is possible to define probability amplitudes, and hence a probability distributions, for $q$ and $p$ separately.  In Ref.~\cite{amin_solutions_2025}, a procedure for obtaining $(q,p)$-separable solutions of the KvN equation~\eqref{eq:KvNequation} was shown.  These separable solutions, upon modulus-squaring, produce separable probability distributions where no statistical correlation exists between $q$ and $p$.  In bra-ket notation, a separable vector $|\chi\rangle$ and its $(q,p)$ representation are given by
\begin{align}\label{eq:qpseparable}
	|\chi\rangle &= |\chi^{(1)}\rangle \circledtimes |\chi^{(2)}\rangle~,\\
	\chi(q,p,t) =
	\langle q,p | \chi \rangle &= \langle q | \chi^{(1)} \rangle \, \langle p | \chi^{(2)} \rangle
	= \chi^{(1)}(q,t) \chi^{(2)}(p,t)~.
\end{align}
More general ($(q,p)$-correlated) vectors can be produced from a superposition
\begin{align}\label{eq:qpcorrelated}
	|\chi \rangle = \sum_n c_n \, |\chi_n^{(1)}\rangle \circledtimes |\chi_n^{(2)}\rangle~.
\end{align}

Note that $\chi^{(1)}(q,t)$ contains only partial information, it is not a probability amplitude for variables that depend on the momentum $p$.  This partial probability amplitude is therefore distinct from approaches defining a classical probability amplitude that is capable of providing complete information for both $q$ and $p$.  These approaches, present for example in Refs.~\cite{holland_quantum_1993} and~\cite{schleich_schrodinger_2013, schleich_wave_2015}, are based on the Hamilton-Jacobi equation and, unlike the KvN approach, possess a nonlinear equation of motion.

The \textit{classical assumption}, that Liouville mechanics is a complete theory, will be our guide.  For us, this implies that a formulation of classical mechanics in Hilbert space should reproduce Liouville mechanics, and nothing more.  This necessitates the treatment of both $q$ and $p$ as fundamental, and forces the use of two Hilbert spaces instead of one for every degree of freedom.  This difference between classical and quantum mechanics will help us keep track of the distinctions between the two theories, even when the language is identical.  The second part of the classical assumption, that the probability distribution $\rho_{_L}$ is invariant under canonical transformations, is discussed in the next section.

\section{Canonical transformations}\label{sec:cantran}

The Poisson bracket ${\lc}\cdot, \cdot{\rc}$, appearing in the Liouville equation~\eqref{eq:liouvilleeq}, is a binary operation satisfying the following rules.  For any three functions $u(q,p,t)$, $v(q,p,t)$ and $w(q,p,t)$, and scalars $c$ or $c(t)$, the Poisson bracket obeys
\begin{subequations}\label{eq:poissonbracket}
\begin{align}
	{\lc}c, u{\rc} &= 0~,\\
	{\lc}v, u{\rc} &= - {\lc}u, v{\rc}~,\label{eq:poissonantisymmetry}\\
	{\lc}u + v, w{\rc} &= {\lc}u, w{\rc} + {\lc}v, w{\rc}~,\\
	{\lc}uv, w{\rc} &= {\lc}u, w{\rc} v + u {\lc}v, w{\rc}~, \label{eq:poissonleibniz}\\
	{\lc} {\lc}u, v{\rc}, w{\rc} & = {\lc} {\lc}u, w{\rc}, v{\rc} + {\lc}u, {\lc}v, w{\rc} {\rc}~.
\end{align}
\end{subequations}
The fundamental Poisson relations between the two fundamental variables $q$ and $p$
\begin{align}\label{eq:fundamentalpoisson}
	{\lc}q, p{\rc} = 1
\end{align}
allows for the calculation of the Poisson bracket of any two functions of them (and of time) as
\begin{align}\label{eq:poissonuv}
	{\lc}u, v{\rc} = \partial_q u \partial_p v - \partial_p u \partial_q v~.
\end{align}
We call variables $(q,p)$ obeying~\eqref{eq:fundamentalpoisson} canonical Poisson conjugates, to differentiate them from canonical commutator conjugates, which will appear in Sec.~\ref{sec:canrep}.  To be precise, if ${\lc}q,p{\rc} = 1$ then we say that the Poisson conjugate of $q$ is $p$, while the Poisson conjugate of $p$ is $-q$, since ${\lc}p, -q{\rc} = 1$.

Any two variables $q'$ and $p'$ obeying the same fundamental relation ${\lc}q', p'{\rc} = 1$ may be used to describe the system, and we have
\begin{align}\label{eq:primepoissonuv}
	{\lc}u, v{\rc} = \partial_q u \partial_p v - \partial_p u \partial_q v
	= \partial_{q'} u \partial_{p'} v - \partial_{p'} u \partial_{q'} v~.
\end{align}
Canonical transformations $(q,p) \to \big(q'(q,p,t),p'(q,p,t) \big)$ are those that preserve the Poisson bracket: ${\lc}u(q',p',t), v(q',p',t){\rc} = {\lc}u(q,p,t), v(q,p,t){\rc}$.  It is well known that the structure of classical Hamiltonian (and, by extension, Liouvillian) mechanics remains invariant under canonical transformations~\cite{balescu_equilibrium_1975, Sudarshan2016}.  In fact, the properties of the Poisson bracket~\eqref{eq:poissonbracket} may be viewed as a consequence of demanding that dynamics remain invariant under certain conditions: if $u$ and $v$ are dynamical variables, then so are $cu$, $u+v$, $uv$, and ${\lc}u,v{\rc}$~\cite{balescu_equilibrium_1975}.  Formulating laws, like Eq.~\eqref{eq:liouvilleeq}, in terms of Poisson brackets expresses their invariance under canonical transformation.  Dynamical evolution is an important example of a continuous canonical transformation.

The Liouville probability distribution, the main character in Liouville mechanics, transforms under a continuous canonical transformation $(q,p) \to (q',p')$ parameterized by $\gamma$ as~\cite{balescu_equilibrium_1975, Sudarshan2016}:
\begin{align}\label{eq:liouvillecantran}
	\begin{split}
	q \to q' = e^{\gamma {\lc}G, \cdot{\rc}} q~, \quad
	p \to p' = e^{\gamma {\lc}G, \cdot{\rc}} p~, \\
	\rho_{_L}(q,p,t) \to \rho'_{_L}(q',p',t) = e^{\gamma {\lc}G, \cdot{\rc}} \rho_{_L}(q,p,t)~.
	\end{split}
\end{align}
Here $G(q,p,t)$ is a real phase-space function that defines the transformation.  An operator ${\lc}f, \cdot{\rc}$, defined by $f(q,p,t)$, acts on a phase-space function $g(q,p,t)$ as
\begin{align}\label{eq:halfpoisson}
	{\lc}f, \cdot{\rc}^n g = {\lc}f, \cdot{\rc}^{n-1} {\lc}f, g{\rc}~ \Rightarrow
	e^{{\lc}f, \cdot{\rc}} g = g + {\lc}f, g{\rc} + \frac{1}{2!}{\lc}f, {\lc}f, g{\rc} {\rc} + \cdots
\end{align}
where $n = 1, 2, \cdots$, and ${\lc}f,\cdot{\rc}^0 = 1$.  The continuous canonical transformation~\eqref{eq:liouvillecantran} takes a form akin to a Taylor expansion employing the differential operator ${\lc}G,\cdot{\rc}$.  Here are three examples
\begin{align}
	\rho_{_L}(q + \Delta q, p, t) &= e^{-\Delta q {\lc}p, \cdot{\rc}} \rho_{_L}(q,p,t) = e^{\Delta q \partial_q} \rho_{_L}(q,p,t)~,\\
	\rho_{_L}(q, p + \Delta p, t) &= e^{\Delta p {\lc}q, \cdot{\rc}} \rho_{_L}(q,p,t) = e^{\Delta p \partial_p} \rho_{_L}(q,p,t)~,\\
	\rho_{_L}(q, p, t + \Delta t) &= e^{\Delta t {\lc}H, \cdot{\rc}} \rho_{_L}(q,p,t) = e^{\Delta t \partial_t} \rho_{_L}(q,p,t)~.
\end{align}
The first two of these are translations in $q$ and $p$, respectively.  The third is a restatement of the Liouville equation of motion~\eqref{eq:liouvilleeq}.  If we define a ``dynamical time'' variable $\tau(q,p)$ such that ${\lc}\tau, H{\rc} = 1$ (see Ref.~\cite{bhamathi_time_2003}), then the third of these examples gives
\begin{align}
	e^{\Delta t {\lc}H, \cdot{\rc}} \rho_{_L}(q,p,t) = e^{-\Delta t \partial_\tau} \rho_{_L}(q,p,t)~,
\end{align}
and the Liouville equation becomes $\partial_t \rho_{_L} + \partial_\tau \rho_{_L} = 0$.

The role of the function $G(q,p,t)$ in the transformation~\eqref{eq:liouvillecantran} is to define the direction in which $\rho_{_L}$ is changing.  As illustrated by the three examples above, the operator ${\lc}G, \cdot{\rc}$ is a translation operator in the direction of the Poisson conjugate of $G$.  To confirm that, perform a canonical transformation $(q,p) \to (G, p_G)$, where $p_G$ is the Poisson conjugate of $G$, i.e., ${\lc}G, p_G{\rc} = 1$.  Now observe that, as seen from~\eqref{eq:primepoissonuv}, ${\lc}G, \cdot{\rc} = \partial_{p_G}$, which effects a translation in the direction of $p_G$.  For that reason, $G$ is sometimes called the generator of the canonical transformation (for example, $p$ as the generator of translations in $-q$), perhaps as a nod to the similar terminology of quantum mechanics.  However, it is crucial to emphasize that it is the operator ${\lc}G, \cdot{\rc}$ that does the work here, not $G$.  This distinction will become clear once we consider the canonical transformation of the probability amplitude $\chi$.

Let $|\chi\rangle$ denote a vector in the Hilbert space $\mathcal{H}$ such that
\begin{align}\label{eq:qpchirho}
	|\langle q,p | \chi \rangle|^2 = \rho_{_L}~.
\end{align}
If a continuous canonical transformation, parameterized by $\gamma$, is effected on $|\chi\rangle$ through an operator $\wh{U}(\gamma)$ as
\begin{align}
	|\chi\rangle \to |\chi'\rangle = \wh{U}(\gamma) |\chi\rangle~,
\end{align}
then $\wh{U}(\gamma)$ must be unitary in order to preserve the norm
\begin{align}
	\langle \chi' | \chi' \rangle =
	\int_{\underline{q}}^{\overline{q}} \int_{\underline{p}}^{\overline{p}} dq' dp' \,
	\rho'_{_L}(q',p',t) = 1 =
	\int_{\underline{q}}^{\overline{q}} \int_{\underline{p}}^{\overline{p}} dq dp \,
	\rho_{_L}(q,p,t) =
	\langle \chi | \chi \rangle~.
\end{align}
Since we are concerned with continuous transformations, the operator $\wh{U}(\gamma)$ can be written as an exponential
\begin{align}
	\wh{U}(\gamma) = e^{-\frac{i}{\hbar} \gamma \wht{G}}~,
\end{align}
where $\wht{G}$ is a Hermitian operator.  The factor of $\frac{1}{\hbar}$ is included to give $\wht{G}$ the dimension of $G$.  We now have the canonical transformation of $|\chi\rangle$ in abstract notation as
\begin{align}\label{eq:generator}
	|\chi\rangle \to |\chi'\rangle =
	e^{-\frac{i}{\hbar} \gamma \wht{G}} |\chi\rangle~.
\end{align}

The operator $-\frac{1}{\hbar}\wht{G}$ appears, in perfect analogy with the usual notation of quantum mechanics, as the generator of the transformation.  Quantum mechanics poses no restriction on its generators of unitary transformations.  By contrast, our set-up here does.  The classical assumption requires correspondence with Liouville mechanics, thus, a canonical transformation of the probability amplitude $\langle q,p | \chi\rangle$ must be derived from the canonical transformation of the probability distribution $\rho_{_L}$.  If the $(q,p)$-representation of $\wht{G}$ is
\begin{align}
	\wt{G} \chi(q,p,t) \coloneqq \langle q,p | \wht{G} | \chi \rangle~,
\end{align}
then the $(q,p)$-representation of Eq.~\eqref{eq:generator} gives
\begin{align}
	\chi^*\chi \to \chi'^*\chi' =
	\left( e^{\frac{i}{\hbar} \gamma \wt{G}} \chi^* \right)
	\left( e^{-\frac{i}{\hbar} \gamma \wt{G}} \chi \right)
\end{align}
which, given that $\chi^*\chi = \rho_{_L}$, must be consistent with the canonical transformation~\eqref{eq:liouvillecantran} of $\rho_{_L}$.  Thus, we must have
\begin{align}\label{eq:chicantran}
	\chi^*\chi + \frac{i}{\hbar} \gamma \left(
	\chi \wt{G} \chi^* - \chi^* \wt{G} \chi \right) + \cdots
	= \rho_{_L} + \gamma {\lc}G, \cdot{\rc} \rho_{_L} + \cdots~.
\end{align}
Equating the coefficients of $\gamma$ on both sides, and using $\rho_{_L} = \chi^*\chi$ and the Leibniz property of the Poisson bracket~\eqref{eq:poissonleibniz}, the first order terms gives (recall that $G$ is real and $\wt{G}$ is Hermitian)
\begin{align}
	\left[ \frac{1}{\chi} \left(
	\frac{1}{i\hbar} \wt{G} - {\lc}G, \cdot{\rc} \right) \chi \right]^* =
	- \frac{1}{\chi} \left(
	\frac{1}{i\hbar} \wt{G} - {\lc}G, \cdot{\rc} \right) \chi~.
\end{align}
This equation is satisfied if
\begin{align}\label{eq:tildeG}
	\wt{G} = i\hbar {\lc}G, \cdot{\rc} + \alpha_G(q,p,t)~,
\end{align}
where $\alpha_G(q,p,t)$ is any real function of $(q,p,t)$, with the same dimension as $G$ and $\wt{G}$.  In summary, if ${\lc}G, \cdot{\rc}$ effects a canonical transformation of $\rho_{_L}$, then $-\frac{1}{\hbar} \wht{G}$ is the generator of the corresponding transformation of $|\chi\rangle$, and its $(q,p)$-representation is given by~\eqref{eq:tildeG}.

The tilde-Hamiltonian $\wt{H}$ in the equation of motion~\eqref{eq:KvNequation} for $\chi(q,p,t)$ is a special instance of~\eqref{eq:tildeG}, and the KvN equation of motion can be derived from
\begin{align}
	\chi(q,p,t + \Delta t) = e^{-\frac{i}{\hbar} \Delta t \wt{H}} \chi(q,p,t)~.
\end{align}
The function $\alpha_H$ associated with $\wt{H}$ is related to the phase of $\chi(q,p,t)$.  The arbitrariness of $\alpha_H$ allows for employing the technique of separation of variables for $q$ and $p$ to solve the eigenvalue equation $\wt{H} \chi_\ve = \ve \chi_\ve$ as detailed in Ref.~\cite{amin_solutions_2025}.  Arbitrary function like $\alpha_G$ are discussed in Sec.~\ref{sec:repgauge}.

Let us examine, in some detail, the effect of a transformation generated by $-\frac{1}{\hbar}\wt{G}$ on $\chi(q,p,t)$.  As discussed above, $e^{\gamma {\lc}G, \cdot{\rc}}\rho_{_L}$ is a canonical transformation of $\rho_{_L}$ that can be viewed as a ``translation'' of $\rho_{_L}$ along $p_G$ by an amount $\gamma$.  For $\chi$, however, the effect of $e^{-\frac{i}{\hbar} \gamma \wt{G}}$ is to first add a phase to $\chi$ then translate the outcome along $p_G$ by an amount $\gamma$.  To see this, recall the Baker-Campbell-Hausdorff formula
\begin{align}
	e^X e^Y = e^Z~; \quad Z = X + Y + \frac{1}{2} [X, Y] +
	\frac{1}{12} \Big( [X, [X,Y]] + [Y, [Y, X]] \Big) + \cdots
\end{align}
and identify $X = \gamma {\lc}G, \cdot{\rc}$, and $Z = -\frac{i}{\hbar} \gamma \wt{G} = \gamma {\lc}G, \cdot{\rc} - \frac{i}{\hbar} \gamma \alpha_G(q,p,t)$.  We then have
\begin{align}\label{eq:alphaBCH}
	-\frac{i}{\hbar} \gamma \alpha_G = Y + \frac{1}{2} [X, Y] +
	\frac{1}{12} \Big( [X, [X,Y]] + [Y, [Y, X]] \Big) + \cdots~.
\end{align}
but $\alpha_G$ is a real function of $(q,p,t)$, then $Y$ must be an imaginary function of the same variables, say, $\frac{i}{\hbar} \gamma \varphi(q,p,t)$, for some real $\varphi$.  This gives the advertised result
\begin{align}
	e^{-\frac{i}{\hbar} \gamma \wt{G}} \chi(q,p,t) =
	e^{\gamma {\lc}G,\cdot{\rc} - \frac{i}{\hbar} \gamma \alpha_G} \chi(q,p,t) =
	e^{\gamma{\lc}G,\cdot{\rc}} \left(
	e^{-\frac{i}{\hbar} \gamma \varphi} \chi(q,p,t) \right)~.
\end{align}
From~\eqref{eq:alphaBCH}, the arbitrary function $\alpha_G$ is related to a phase change of $\chi$ as
\begin{align}
	- \alpha_G = \varphi + \frac{1}{2} {\lc}G, \varphi{\rc} +
	\frac{1}{12} {\lc}G, {\lc}G, \varphi{\rc}{\rc} + \cdots~.
\end{align}
The relation between the arbitrary $\alpha_G$ and the phase of the probability amplitude in the special case of $G = H$ is related to a $U(1)$ gauge symmetry of the KvN equation~\eqref{eq:KvNequation}, and has been studied, for example, in Refs.~\cite{Klein2017, Bondar2019, amin_solutions_2025}.

Finally, for completeness, let us see the condition for the Hermiticity of the operator $\wt{G} = i\hbar {\lc}G,\cdot{\rc} + \alpha_G$.  It is easy to show that, if $G$ is real, then
\begin{align}
	\langle \chi_a | \wt{G} \chi_b \rangle =
	\langle \wt{G} \chi_a | \chi_b \rangle +
	i\hbar \int_{\underline{q}}^{\overline{q}} \int_{\underline{p}}^{\overline{p}} dq dp \,
	{\lc}G, \chi_a \chi_b{\rc}~, \quad
	\forall \, |\chi_a\rangle \, , |\chi_b\rangle \in \mathcal{H}~,
\end{align}
where the inner product $\langle \chi' | \chi \rangle$ is $\int \int dq dp \chi'^* \chi$.  The condition for the Hermiticity of $\wt{G}$ then is for the second term on the right-hand-side to vanish for all vectors $|\chi_a\rangle$ and $|\chi_b\rangle$ in the space $\mathcal{H}$.

The condition that unitary transformations of $\chi$ correspond to canonical transformations of $\rho_{_L}$ is to ensure that our Hilbert-space formulation corresponds to classical mechanics.  No such restriction exists for quantum mechanics.  There, no prior conditions are placed on the probability distribution that dictates the allowed transformations of the probability amplitude.  The classically allowed Hermitian operators, which we denote \textit{tilde-variables}, and their properties are discussed in the next section.

\section{The tilde-variables}\label{sec:tv}

For every phase-space variable $u(q,p,t)$, there is an operator $\wht{u}$ that generates a canonical transformation of $|\chi\rangle$ as a translation in the opposite direction of the Poisson conjugate of $u$.  This was detailed in the previous section.  These operators, which we call tilde-variables, are part of the structure of classical mechanics that were uncovered in a Hilbert-space formulation.  We now turn to their algebraic and representation properties.  In this section, we will focus on the $(q,p)$-representation of $\wht{u}$
\begin{align}\label{eq:utilde}
	\langle q,p | \wht{u} | \chi \rangle = \wt{u} \chi~, \quad
	\wt{u} = i\hbar {\lc}u, \cdot{\rc} + \alpha_u~.
\end{align}
We say this is the ``$(q,p)$''-representation, though the representation above is valid if we choose any other pair of Poisson conjugates, like $(u,p_u)$ for instance; $q$ and $p$ here should be interpreted as any such pair obeying ${\lc}q,p{\rc}=1$.

It is easy to show, given the definition~\eqref{eq:utilde} of tilde variables, that for any two phase-space variables $u(q,p,t)$ and $v(q,p,t)$ we have
\begin{align}\label{eq:poissoncommutator}
	\frac{1}{i\hbar} [u, \wt{v}] =
	\frac{1}{i\hbar} [\wt{u}, v] =
	{\lc}u, v{\rc}~.
\end{align}
This relation provides a map connecting the Poisson bracket of any two phase-space variables to a commutator between phase-space- and tilde-variables.  Therefore, given any classical theory formulated in phase-space, a direct translation of it into Hilbert-space language exists.

The tilde-position $\wt{q}$ and tilde-momentum $\wt{p}$, special cases of tilde-variables obeying $[q, \wt{p}] = [\wt{q}, p] = i\hbar$, are known in the literature, see for example Refs.~\cite{Mauro2002, Gozzi2004, Bondar2012, Wilczek2015, amin_solutions_2025}.\footnote{
	In Refs.~\cite{Mauro2002, Gozzi2004, Bondar2012, Wilczek2015}, and others, operators $\lambda_q := -i \partial_q$ and $\lambda_p := -i \partial_p$ appear as commutator conjugates of $q$ and $p$ respectively.  In our notation, $\lambda_q = \frac{1}{\hbar} (\wt{p} - \alpha_p)$ and $\lambda_p = -\frac{1}{\hbar} (\wt{q} - \alpha_q)$.
}
However, their roles unclear, they are often considered ``extra'' or ``auxiliary'' variables.  Here, we see that they are, like other tilde-variables, generators of canonical transformations.

A crucial difference between classical and quantum mechanics is elucidated upon examining the generators of position and momentum translations.  In QM, if the position $q_Q$ is taken to be fundamental, then the quantum momentum $p_Q$ is defined as the generator of space translations (up to some constant factor)~\cite{ballentine_quantum_2014}.  Similarly, if $p_Q$ is taken to be the fundamental quantity, then $-q_Q$ can be defined as the generator of momentum translations.  In the classical case, as we have seen, both $q$ and $p$ are fundamental, and the generators of their translations (up to a constant factor) are $\wt{p}$ and $-\wt{q}$ respectively.  This suggests the possibility that the quantum momentum is the analogue of $\wt{p}$, rather than $p$.  More explorations of this theme are to follow.

\subsection{Algebra}\label{sec:algebra}

First, we turn to some properties of the tilde-variables (generators of canonical transformations).  We construct a tilde-variable given two known ones $\wt{u}$ and $\wt{v}$, and scalars $c(t)$, using simple binary operations.  From~\eqref{eq:utilde} we can prove the following for the basic operations 
\begin{subequations}\label{eq:tildeimages}
\begin{align}
	\wt{c} &= \alpha_c~,\\
	\wt{cu} &= c\wt{u} + \alpha_{cu} - c\alpha_u~,\\
	\wt{u+v} &= \wt{u} + \wt{v} + \alpha_{u+v} - \alpha_u - \alpha_v~,\\
	\wt{uv} &= u\wt{v} + v\wt{u} + \alpha_{uv} - u\alpha_v - v\alpha_u~,\\
	\wt{{\lc}u,v{\rc}} &= \frac{1}{i\hbar} [\wt{u}, \wt{v}] +
	\alpha_{{\lc}u, v{\rc}} - {\lc}u, \alpha_v{\rc} - {\lc}\alpha_u, v{\rc}~.\label{eq:tildepoisson}
\end{align}
\end{subequations}
These construct tilde-variables from phase-space ones.  They are somewhat simplified by adopting the canonical representation, discussed in Sec.~\ref{sec:canrep} (Eq.~\eqref{eq:canrep}).

The above equations show that the scalar multiplication, sum, and commutator of tilde-variables is a tilde-variable.  Note, however, that the product of two tilde-variables is not a tilde-variable.  That is, a function $f(\wt{u})$ will not generate a canonical transformation, unless it is linear in $\wt{u}$.  All Hermitian operators that we deal with here are going to be linear in tilde-variables.  Once more, this is a point of difference between classical and quantum mechanics in Hilbert space; the latter puts no restrictions on the Hermitian operators appearing in the theory.  It stems, as discussed in the paragraph containing Eq.~\eqref{eq:chicantran}, from the classical requirement that transformations of $\chi$ must derive from canonical transformations of $|\chi|^2 = \rho_{_L}$.

\subsection{Representation gauge}\label{sec:repgauge}

The presence of arbitrary functions in $\wt{u}$ expresses the nonuniqueness of the $(q,p)$-representation of $\wht{u}$.  We have shown in Sec.~\ref{sec:cantran} that, for $\wt{u}$, the arbitrary function $\alpha_u$ is related to the phase of $\chi$.  This tells us that the nonuniqueness of the representation of $\wht{u}$ is associated with the nonuniqueness of the correspondence of $\chi$ to $\rho_{_L}$ (an arbitrary phase can be added to the former without affecting the latter).  We may call a particular choice of $\alpha_u$ a ``representation gauge'' since, in the case of $u=H$, it is related to a $U(1)$ gauge symmetry.

To see the effect of this nonuniqueness, consider the eigenfunctions of $\wt{u}$.  These are solutions of the eigenvalue equation
\begin{align}
	\wt{u} \chi_\lambda = \lambda \chi_\lambda~.
\end{align}
If we consider $\chi_\lambda$ as a function of $u$ and its Poisson conjugate $p_u$, then $\wt{u} = i\hbar \partial_{p_u} + \alpha_u$ and its eigenfunctions can be expressed formally as
\begin{align}
	\chi_\lambda = f_\lambda(u,t) \, e^{-\frac{i}{\hbar} \left[
		\lambda p_u - \int \alpha_u dp_u \right]}~,
\end{align}
where $f_\lambda(u,t)$ is any normalizable function.  We see that different choices of $\alpha_u$ change the phase of $\chi_\lambda$ and, consequently, change the eigenvalue $\lambda$ itself.  We may say that $\wt{u}$ is gauge- or representation-dependent.

This apparent unreliability of the eigenvalues of tilde-variables may seem unacceptable.  However, observe that the same representation ambiguity/redundancy exists in quantum mechanics.  If the quantum mechanical momentum $\wh{p}_Q$ is defined, in relation to position $\wh{q}_Q$, through the fundamental commutator relation
\begin{align}
	\frac{1}{i\hbar} [\wh{q}_Q, \wh{p}_Q] = 1~,
\end{align}
then the position-representation of the momentum operator contains an arbitrary function of $q_Q$
\begin{align}\label{eq:QMp}
	p_Q = -i\hbar \partial_{q_Q} + f(q_Q)~.
\end{align}
This representation ambiguity of the momentum operator is noted in quantum mechanics textbooks, see Ref.~\cite[p. 210]{shankar_principles_1994}, for example.  The quantum mechanical momentum, too, possesses a representation-dependent spectrum.  The representation ambiguity is often dealt-with by silently setting $f(q_Q) = 0$ in~\eqref{eq:QMp} and is usually ignored by force of habit, but comes to attention when encountered in the relatively unfamiliar context of Hilbert-space classical mechanics.

\subsection{Canonical representation}\label{sec:canrep}

Let us now consider a reasonable fixing of representation, i.e., setting the arbitrary function $\alpha_u$ for all tilde-variables $\wt{u}$.  For that, we adopt two guiding rules.  The first is to pick a Poisson pair, say $q$ and $p$, and to express all $\alpha$'s in terms of the pair's arbitrary functions $\alpha_q$ and $\alpha_p$.  This reduces the infinite number of arbitrary functions down to two (for a system with a single degree of freedom).

Inspecting Eqs.~\eqref{eq:tildeimages}, we may impose choices of the $\alpha$'s that simplifies the construction of tilde-variables as
\begin{subequations}\label{eq:canrep}
\begin{align}
	\alpha_c = 0~ &\Rightarrow \wt{c} = 0~,\label{eq:tildeofc}\\
	\alpha_{cu} = c\alpha_u &\Rightarrow \wt{cu} = c\wt{u}~,\\
	\alpha_{u+v} = \alpha_u + \alpha_v &\Rightarrow \wt{u+v} = \wt{u} + \wt{v}~,\\
	\alpha_{uv} = u\alpha_v + v\alpha_u &\Rightarrow \wt{uv} = u\wt{v} + v\wt{u}~,\\
	\alpha_{{\lc}u, v{\rc}} = {\lc}u, \alpha_v{\rc} + {\lc}\alpha_u, v{\rc} &\Rightarrow
	\wt{{\lc}u,v{\rc}} = \frac{1}{i\hbar} [\wt{u}, \wt{v}]~.\label{eq:tildepoissoncr}
\end{align}
\end{subequations}
The first four of these comply with the first rule.  In particular, they imply that for any $u(q,p,t)$ analytical in $q$ and $p$,  $\wt{u}$ can be expressed in terms of $\wt{q}$ and $\wt{p}$ as
\begin{align}
	\begin{split}
	\wt{u}
	&= i\hbar {\lc}u, \cdot{\rc} + \alpha_u\\
	&= i\hbar {\lc}u, \cdot{\rc} + \partial_q u \alpha_q + \partial_p u \alpha_p\\
	&= \partial_q u \wt{q} + \partial_p u \wt{p}~.
	\end{split}
\end{align}
More generally, we get
\begin{align}
	\wt{f(u)} = f'(u) \, \wt{u}~,
\end{align}
where the prime denotes total derivative with respect to the single argument, for an analytical function $f(u)$.

The second guiding rule is to set a vanishing commutator between between $\wt{q}$ and $\wt{p}$.  This is achieved through the first and the last equations in~\eqref{eq:canrep}
\begin{align}\label{eq:qpcanrep}
	\frac{1}{i\hbar} [\wt{q}, \wt{p}] = \wt{{\lc}q,p{\rc}} = \wt{1} = 0~.
\end{align}
Besides simplifying calculations, the choices~\eqref{eq:canrep} of the $\alpha$'s allow us to think of $(q, \wt{p})$ and $(\wt{q}, p)$ as two independent pairs of canonical commutator (as opposed to Poisson) conjugates
\begin{subequations}\label{eq:fundamentalcommutator}
\begin{align}
	&\frac{1}{i\hbar} [q, \wt{p}] =
	\frac{1}{i\hbar} [\wt{q}, p] = 1~,\\ 
	&\frac{1}{i\hbar} [q, p] =
	\frac{1}{i\hbar} [\wt{q}, \wt{p}] =
	\frac{1}{i\hbar} [q, \wt{q}] =
	\frac{1}{i\hbar} [\wt{p}, p] = 0~.
\end{align}
\end{subequations}
For this reason, we call Eqs.~\eqref{eq:canrep} a \textit{canonical representation} of the tilde variables.  Apart from the condition $\partial_q \alpha_q + \partial_p \alpha_p = 0$ from~\eqref{eq:tildeofc} and~\eqref{eq:tildepoissoncr}, $\alpha_q(q,p,t)$ and $\alpha_p(q,p,t)$ are left free.  The canonical representation will be useful for the Wigner representation discussed in Sec.~\ref{sec:wigner}.

\section{The postulates}\label{sec:postulates}

To formalize the difference between classical and quantum mechanics, it would be helpful to trace their divergence on the postulates level.  Instead of categorizing a theory as classical or nonclassical based on the presence of one effect or the other, we will rely on an axiomatic approach where a theory is defined to be classical or quantum based on its postulates.  If a certain effect can be derived from a classical theory, then it is not characteristic of quantum mechanics.  If mechanics is defined in terms of its variables, states, and their transformations, then classical and quantum mechanics are contrasted solely based on the properties these three concepts.  We will state here four postulates, the first two (copied, verbatim, from Ballentine's QM textbook~\cite{ballentine_quantum_2014}) are common to both CM and QM, while the latter pair sets them apart.

\indent \textbf{Postulate 1}
\textit{To each dynamical variable there is a Hermitian operator whose eigenvalues are the possible values of the dynamical variable.}

According to which, phase-space variables like position and momentum correspond to operators $\wh{q}$ and $\wh{p}$.  In the $(q,p)$-representation, the effect of these operators is multiplicative
\begin{align}
	\langle q,p | \, \wh{q} \, | \chi \rangle = q \chi(q,p,t)~, \quad
	\langle q,p | \, \wh{p} \, | \chi \rangle = p \chi(q,p,t)~.
\end{align}
In that representation, we may write
\begin{align}
	\wh{q} \to q~, \quad \wh{p} \to p~, \quad
	\wh{u} \to u(q,p,t)~,
\end{align}
where $\wh{u}$ is the operator representing any phase-space variable $u$.  This is similar to the position representation in quantum mechanics, except that here it is simultaneously a position and a momentum representation.

What about tilde-variables, the generators of canonical transformation, should they too be considered dynamical variables?  If a tilde-variable is Hermitian, then the completeness of its eigenstates allows for expanding any state in terms of those eigenstates.  An observation that determines the Liouville distribution, for instance, allows us to infer the probability amplitude, and express that as a linear combination of the eigenstates of some Hermitian tilde-variable.  Such expansion then determines a range of eigenvalues of the tilde-variable consistent with the state.  Thus, the tilde-variables are, in principle, observable.  The argument for the observability of variables corresponding to Hermitian operators (expressed differently) is found in QM textbooks~\cite{dirac_principles_1958, ballentine_quantum_2014}, and it is general enough to apply to the classical formulation at hand.

We see then that while tilde-variables are considered dynamical variables, information about them can be obtained through $\chi$ which itself can be deduced from measurements involving phase-space variables only.  This complies with our requirement that the present Hilbert-space formulation, despite its enlargement of the number of dynamical variables through the inclusion of tilde-variables, contains no effects beyond that of Liouville mechanics.

In any case, it suffices for our purposes to see $\wht{q}$, $\wht{p}$, and in general $\wht{u}$, as similar in character to the quantum mechanical momentum operator and functions of it: they are generators of transformations and are considered dynamical variables in their own right.  For completeness, we write the $(q,p)$-representation of tilde-variables
\begin{align}
		\wht{q} \to \wt{q} = i\hbar \partial_p + \alpha_q~, \quad
		\wht{p} \to \wt{p} = -i\hbar \partial_q + \alpha_p~, \quad
		\wht{u} \to \wt{u} = i\hbar{\lc}u, \cdot{\rc} + \alpha_u~,
\end{align}
as discussed at length in Sec.~\ref{sec:cantran}.

\textbf{Postulate 2}
\textit{To each state there corresponds a unique state operator, which must be Hermitian, nonnegative, and of unit trace.}

From the outset, we were concerned with ensembles of classical systems, described by $\rho_{_L}$, a statistical distribution.  This does not exclude ordinary classical mechanics as it is described by a special case of statistical distributions (a Dirac delta function on phase space).  In Hilbert space, a state operator obeying the conditions of postulate 2 is associated with a given ensemble~\cite{ballentine_quantum_2014}.  Specifically, if $\wh{\rho}$ denotes the state operator, then averages of measurements of some dynamical variable $\wh{R}$ (including phase-space as well as tilde-variables) is given by
\begin{align}\label{eq:averagewithtrace}
	\langle \wh{R} \rangle = \Tr (\wh{\rho} \, \wh{R})~,
\end{align}
where Tr denotes the trace.  The consistency of this statistical role of $\wh{\rho}$ requires the conditions listed in postulate 2
\begin{align}\label{eq:stateoperator}
	\wh{\rho}\,^\dagger = \wh{\rho}~, \quad
	\langle \chi | \, \wh{\rho} \, | \chi \rangle \geq 0~~ \forall \, |\chi\rangle~, \quad
	\Tr \wh{\rho} = 1~.
\end{align}
The first of these ensures that the average of a real-valued variable is real, the second that the average of a nonnegative-valued one is nonnegative, and the third that probabilities sum to unity~\cite{ballentine_quantum_2014}.

We may distinguish between two types of state operators: pure and nonpure.  A pure state has the following equivalent properties
\begin{align}\label{eq:pure}
	\wh{\rho} = |\chi\rangle \langle\chi|~, \quad
	\wh{\rho}\,^2 = \wh{\rho}~, \quad
	\Tr \wh{\rho}\,^2 = 1~.
\end{align}
The first of these is the statement that a pure state operator can be written in the form of a ket-bra outer product.  A nonpure state operator is one that can be expressed as convex sum of pure state operators, i.e.,
\begin{align}\label{eq:nonpure}
	\wh{\rho} = \sum_n w_n \wh{\rho}_n~, \quad
	w_n \geq 0~, \quad
	\sum_n w_n = 1~.
\end{align}
where $\wh{\rho}_n$ are pure.  This also implies that $\Tr \wh{\rho}\,^2 \leq 1$, with the equality holding only if $\wh{\rho}$ is pure.  For any two state operators $\wh{\rho}_a$ and $\wh{\rho}_b$, the following inequality can be derived~\cite{ballentine_quantum_2014}
\begin{align}\label{eq:rhoabinequality}
	0 \leq \Tr \big( \wh{\rho}_a \wh{\rho}_b \big) \leq 1~,
\end{align}
with the equality on the right holding only if $\wh{\rho}_a = \wh{\rho}_b$ is a pure state.  This inequality is instrumental for proving the Wigner negativity, discussed in Sec.~\ref{sec:wigner}.

We see that the diagonal elements of the density matrix of a pure state in the $(q,p)$-representation give the Liouville distribution
\begin{align}\label{eq:rho_Lfromrho}
	\langle q,p | \, \wh{\rho} \, | q,p \rangle = |\chi(q,p,t)|^2 = \rho_{_L}(q,p,t)~.
\end{align}
For a nonpure state, we get the convex sum $\sum_n w_n \rho_{_L n}$ of $n$ Liouville distributions, which is itself a Liouville distribution (real, nonnegative, normalized, and obeying the Liouville equation~\eqref{eq:liouvilleeq}).  Finally, if the operator $\wh{R}$ is a phase-space variable, then its average~\eqref{eq:averagewithtrace} in the $(q,p)$-representation is
\begin{align}\label{eq:averagewithliouville}
	\langle \wh{R}(q,p,t) \rangle = \Tr (\wh{\rho} \, \wh{R}) =
	\int_{\underline{q}}^{\overline{q}} \int_{\underline{p}}^{\overline{p}} dq dp \,
	\rho_{_L}(q,p,t) \, R(q,p,t)
\end{align}
as expected.  However, the average of an operator containing tilde-variables cannot be calculated similarly using the Liouville distribution $\rho_{_L}$.  Instead, a Wigner representation, shown in Sec.~\ref{sec:wigner} is needed.

Pure and nonpure (or mixed) states are extensively studied in quantum mechanics.  In classical mechanics, however, they haven't received much attention.  It is sometimes argued that a pure classical state is one that is a delta function distribution on phase space (a single point $(q,p)$), and that all other distributions are nonpure.  This, seemingly, is in analogy with the notion that a pure quantum state represents complete knowledge of a particle, while nonpure state represents less than complete knowledge of the state of the particle.  We will not contest this assumption, but consider it unnecessary.

A different definition of a classical pure state is provided in Holland's book~\cite{holland_quantum_1993}.  There, a pure state is one whose Liouville distribution takes the form
\begin{align}
	\rho_{_L}(q,p,t) = P(q, t) \, \delta \big(p - \partial_q S(q,t) \big)~,
\end{align}
where $P(q,t)$ is a probability distribution on configuration space, and $S(q,t)$ is Hamilton's principal function.  Nonpure states, in Holland's view, can be expressed as a convex combination of pure ones.  In this expression, $P(q,t)$ is not a marginal of the phase-space distribution (that would make it incapable of providing the average of functions of momentum).  Rather, it is a probability distribution for all variables of the system, provided that the momentum is given as a function of position as $p = \partial_q S(q,t)$.  The configuration-space approach, based on the Hamilton-Jacobi equation, is interesting but it is not the focus of this paper.  It is instructive, however, to see definitions of pure and nonpure states that go beyond the simple one discussed in the previous paragraph.

The definition of pure and nonpure states we use in Eqs.~\eqref{eq:pure} and~\eqref{eq:nonpure} is identical to the quantum definition (as mentioned above, a big part of this section is based on Ballentine's QM textbook~\cite{ballentine_quantum_2014}).  It places the emphasis on the state operators, not the (Liouville) probability distribution.  In doing so, it adopts the same language as that of QM and eliminates superficial differences between the two theories.  Conceptually, we may view a pure state as one corresponding to a definite ensemble, while a nonpure state as one where there is a ``statistical mixture'' of ensembles.  This understanding is valid in both the classical and the quantum case.

We now write the equation of motion for the state operator
\begin{align}\label{eq:rhoeom}
	\frac{d}{dt} \wh{\rho} + \frac{1}{i\hbar} [\wh{\rho}, \wht{H}] = 0~.
\end{align}
This is identical to the quantum von Neumann equation of motion, but where the operator responsible for time-evolution is $\wht{H}$, as discussed in Sec.~\ref{sec:probamp}.  As expected, sandwiching this equation between bra $\langle q,p|$ and ket $|q,p \rangle$, and using the completeness~\eqref{eq:qpcompleteness} of the $|q,p\rangle$ basis and the Poisson-commutator map~\eqref{eq:poissoncommutator}, produces the Liouville equation~\eqref{eq:liouvilleeq}.

So far in this section, we have been implicitly employing the classical assumption, discussed in previous sections, in that we assumed the commutativity of the position $\wh{q}$ and the momentum $\wh{p}$ operators.  We now formalize this into one of two classical postulates.

\textbf{Postulate 3}
\textit{The classical Hilbert space $\mathcal{H}$ is a tensor product of $2n$ Hilbert spaces corresponding to $n$ degrees of freedom and their Poisson conjugates.}

With this postulate, we encounter the first major difference between classical and quantum mechanics.  The quantum analogue of this postulate is that every degree of freedom gets only one Hilbert space.  It is justified, as argued in Sec.~\ref{sec:probamp}, by the experimental fact that the prediction of motion for a single degree of freedom requires two independent pieces of information, e.g., position and momentum.  This also implies the commutativity of $\wh{q}$ and $\wh{p}$.

Let $\mathcal{H}_1$ denote the Hilbert space spanned by $|q\rangle$, and $\mathcal{H}_2$ that spanned by $|p\rangle$.  In analogy with QM, where we may choose to span the space with position eigenvectors or the eigenvectors of their associated generator of translations (quantum momentum), we may use $|\wt{p}\rangle$ to span $\mathcal{H}_1$ or $|\wt{q}\rangle$ to span $\mathcal{H}_2$.  Marginal probability amplitudes over $q$ and $p$ were already discussed towards the end of Sec.~\ref{sec:probamp}.  State operators pertaining to $\mathcal{H}_1$ or $\mathcal{H}_2$ can be defined by taking partial traces of $\wh{\rho}$
\begin{align}
	\wh{\rho}\,^{(1)} \coloneqq \Tr^{(2)} \wh{\rho}
	= \int_{\underline{\wt{q}}}^{\overline{\wt{q}}} d\wt{q} \,
	\langle \wt{q}| \, \wh{\rho} \, |\wt{q} \rangle
	= \int_{\underline{p}}^{\overline{p}} dp \,
	\langle p| \, \wh{\rho} \, |p \rangle~,\\
	\wh{\rho}\,^{(2)} \coloneqq \Tr^{(1)} \wh{\rho}
	= \int_{\underline{q}}^{\overline{q}} dq \,
	\langle q| \, \wh{\rho} \, |q \rangle
	= \int_{\underline{\wt{p}}}^{\overline{\wt{p}}} d\wt{p} \,
	\langle \wt{p}| \, \wh{\rho} \, |\wt{p} \rangle~,
\end{align}
where $(\underline{q}, \overline{q})$, $(\underline{\wt{p}}, \overline{\wt{p}})$, $(\underline{\wt{q}}, \overline{\wt{q}})$, and $(\underline{p}, \overline{p})$ denote the lower and upper limits on $q$, $\wt{p}$, $\wt{q}$, and $p$, respectively.  Partial traces are state operators in their own right, obeying conditions~\eqref{eq:stateoperator}~\cite{ballentine_quantum_2014}.

Despite $\mathcal{H}_1$ and $\mathcal{H}_2$ being separate spaces, tilde-variables in general take a form that couples variables of the two space.  For example, if the Hamiltonian is given by $H = K(p) + U(q)$, then the Liouvillian (in the canonical representation) is
\begin{align}
	\wt{H} = \wt{p} \circledtimes K'(p) + U'(q) \circledtimes \wt{q}~,
\end{align}
where the prime denotes total derivative with respect to the single argument.  The $\mathcal{H}_1$ and $\mathcal{H}_2$ variables are coupled in the tilde-Hamiltonian $\wt{H}$ even when the Hamiltonian $H$ is separable in $q$ and $p$.

\textbf{Postulate 4}
\textit{The classically allowed transformations of the state are canonical transformations.}

Observe that tilde-variables generate canonical transformations, but their products do not, as discussed in Sec.~\ref{sec:algebra}.  The above postulate then implies that transformations generated by products of tilde-variables are not allowed.

The quantum version of this postulate is simply that there is no such restriction.  For example, the square of the generator of space translations (quantum momentum) is itself a valid generator of time translations (as in the case of the quantum free-Hamiltonian).

Note an important consequence of this for the behaviour of probability distributions.  The classical probability distribution $\rho_{_L}$, forced to transform only through canonical transformations, always has a fixed region of support on its underlying space (phase space).  In fact, this is another definition of canonical transformations: they are those that preserve the phase-space area (or volume, for more than one degree of freedom).  By contrast, the quantum mechanical probability distribution (the modulus-square of the wavefunction) may spread over its underlying space (e.g., configuration space) with no restriction on its region of support.

Postulates 3 and 4 may seem inelegant.  They make reference to Poisson brackets, explicitly in postulate 3 and implicitly (through reference to canonical transformations) in postulate 4, which makes classical mechanics seem less native to Hilbert space.  It may be possible to reformulate these postulates in a more self-contained manner.  For example, postulate 3 may be stated as requiring two Hilbert spaces per degree of freedom, without reference to Poisson conjugates.  This will make it almost at home even with QM, provided one associates the extra Hilbert spaces with extra degrees of freedom.  Indeed, Wilczek compared KvN to QM with double the degrees of freedom~\cite{Wilczek2015}.

The fourth postulate, however, is pure restriction.  Even if it were formulated without reference to phase-space concepts, it still feels somewhat out of place among more abstract postulates, and it is.  Nonetheless, it is a restriction that ensures that this very quantum-looking formulation is actually classical.  A more elegant formulation of the postulates may be the subject of future work, but here our main concern is to write classical mechanics in as much as possible the same language as quantum mechanics while maintaining its tether to the ordinary formulation.

\section{Uncertainty relations}\label{sec:uncertainty}

Now we are in position to state the first of two main results of this paper.  After the groundwork laid above, it should come as no surprise that classical mechanics in Hilbert space possesses uncertainty relations.  We have seen that noncommutative variables exist in CM, uncertainty relations follow for any pair of such variables using the standard derivation~\cite{ballentine_quantum_2014, shankar_principles_1994}.  We will focus here on some conceptual aspects of such relations.

Of course, since by assumption position $q$ and momentum $p$ (or any two phase-space variables) commute, there are no nontrivial uncertainty relations connecting them.  For two noncommuting variables, like the position $q$ and tilde-momentum $\wt{p}$ (we drop the hats in this section), we have
\begin{align}
	\sigma_q \sigma_{\wt{p}} \geq \frac{1}{2}
	\big| \langle [q, \wt{p}] \rangle \big|
	= \frac{\hbar}{2}~.
\end{align}
Here, $\sigma_{\wt{p}}$ (and similarly for $\sigma_q$) is the standard deviation given by
\begin{align}
	\sigma_{\wt{p}} \coloneqq \sqrt{\big\langle
	\big( \wt{p} - \langle \wt{p} \rangle \big)^2
	\big\rangle}~.
\end{align}
More generally, using the Poisson-commutator map~\eqref{eq:poissoncommutator}, the uncertainty relation between a phase-space variable and a tilde-variable is
\begin{align}\label{eq:pvtvuncertainty}
	\sigma_u \sigma_{\wt{v}} \geq \frac{\hbar}{2} \,
	\big| \langle {\lc}u, v{\rc} \rangle \big|~.
\end{align}
Between two tilde-variables, using the relation~\eqref{eq:tildepoisson}, the uncertainty relations is
\begin{align}
	\sigma_{\wt{u}} \sigma_{\wt{v}} \geq \frac{\hbar}{2} \,
	\big| \langle \wt{{\lc}u, v{\rc}} \rangle - \langle \alpha_{{\lc}u, v{\rc}} - {\lc}u, \alpha_v{\rc} - {\lc}\alpha_u, v{\rc} \rangle \big|~.
\end{align}
The last three terms on the right-hand-side disappear if we use the canonical representation of tilde-variables, according to~\eqref{eq:tildepoissoncr}.

It was argued in Ref.~\cite{bermudez_manjarres_three_2024} that such uncertainty relations cannot exist, because their calculation depends on the square of a tilde-variable and, as we have discussed in Sec.~\ref{sec:tv}, the square of a tilde-variable is not itself a tilde-variable (i.e., is not a generator of canonical transformation).\footnote{
	In Ref.~\cite{bermudez_manjarres_three_2024}, the authors use variables they call the van Hove operators.  These, in our scheme, are particular representations of tilde-variables: they correspond to particular choices of the arbitrary $\alpha$'s discussed in Sec.~\ref{sec:tv}.
}
This argument, however, denies the existence of a standard deviation for tilde-variables altogether, which is implausible.  An expectation value of a squared tilde-variable $\langle \chi | \wt{u}^2 | \chi \rangle$ may simply be calculated as the inner product $\langle \wt{u}^\dagger \chi | \wt{u} \chi \rangle$, which involves only the effect of $\wt{u}$, not $\wt{u}^2$, on $|\chi\rangle$.  Further, and more fundamentally, we cannot deny the existence of mathematical objects like $\wt{u}^2$, we merely require that they are not used to generate transformations of physical significance on the state.  For example, time-evolution cannot be generated by such operators.  There is no restriction on their use for performing calculations, much like the use of the unobservable (non-Hermitian) ladder operators in quantum mechanics.

The classical uncertainty relations are not mysterious or befuddling, they can be completely explained in terms of phase-space Liouville mechanics.  The Hilbert-space formulation merely provided a mechanism to generate them.  Just as in quantum mechanics, there cannot be a state that corresponds to a definite value of two noncommuting operators, and less uncertainty in one implies more in the other.  For example, the more localized an ensemble is in position $q$, the more spread it is in $\wt{p}$, and similarly for $\wt{q}$ and $p$.  An example of the uncertainty trade-off between the dynamical time $\tau$  and the tilde-Hamiltonian $\wt{H}$ was studied in Ref.~\cite{amin_solutions_2025}.  Since $\tau$ and $H$ are Poisson conjugates (${\lc}\tau, H{\rc}=1$) then, from~\eqref{eq:poissoncommutator}, $\tau$ and $\wt{H}$ are commutator conjugates ($\frac{1}{i\hbar} [\tau, \wt{H}] = 1$), implying an uncertainty relation between them.  Since the dynamical time $\tau(q,p)$ is a phase-space variable tracking the time evolution of the system (see Ref.~\cite{bhamathi_time_2003}), it is easy to see that an ensemble defined by having a definite dynamical time cannot be an equilibrium ensemble (one that does not evolve in time).

It can be seen then that uncertainty relations exist in CM in the same way they do in QM: an uncertainty trade-off between a variable and its associated generator of transformations.  The key difference, then, between CM and QM is not the existence of uncertainty, but in the physical meaning ascribed to the generators.  We discussed above that QM identifies momentum with the generator of space translation, while CM does not (it is $\wt{p}$, not $p$, that generates the translation).  We do not attempt to explore these fundamental differences, but to identify them.

\section{The Wigner representation}\label{sec:wigner}

Now we are in the peculiar position of having started in phase space, migrated to Hilbert space, and then returning back to phase space with some extra variables we picked up along the way.  In translating Liouville mechanics into Koopman-von Neumann mechanics, we learned that tilde-variables exist as dynamical variables, along with their commutation relations with ordinary phase-space variables.  It is well-known that the Wigner representation of Hilbert-space quantum mechanics produces a quasi-probability distribution on phase space.  (For background, see Refs.~\cite{Hancock2004, Zachos2005, Case2008, Curtright2014, ballentine_quantum_2014}).  What would be the Wigner representation of the classical state operator $\wh{\rho}~$?

According to postulate 3 from Sec.~\ref{sec:postulates}, the Hilbert space $\mathcal{H}$ for a single degree of freedom is a tensor product of two spaces $\mathcal{H}_1 \circledtimes \mathcal{H}_2$.  The variables $\wh{q}$ and $\wht{p}$ belong to $\mathcal{H}_1$, while $\wht{q}$ and $\wh{p}$ to $\mathcal{H}_2$.  We will make use of the canonical representation~\eqref{eq:canrep} so that $(\wh{q}, \wht{p})$ and $(\wht{q}, \wh{p})$ are treated as two independent pairs of canonical commutator conjugates, as seen in~\eqref{eq:fundamentalcommutator}.  In the Wigner representation, these operators are represented by the symbols $(q,\wt{p})$ and $(\wt{q},p)$, which form the $\textit{doubled phase space}$ of the system.

The Wigner representation of an operator $\wh{R}$ is given by~\cite{ballentine_quantum_2014}
\begin{align}\label{eq:wignerrep}
	\begin{split}
	R_W(q,\wt{p},\wt{q},p,t) &\coloneqq W[\wh{R}]\\
	&\coloneqq
	\iint_{-\infty}^{\infty} dq' dp' \,
	\big\langle q - q'/2, p - p'/2 \big| \, \wh{R} \, \big| q + q'/2, p + p'/2 \big\rangle
	\, e^{\frac{i}{\hbar} (\wt{p} q' - \wt{q} p')}~.
	\end{split}
\end{align}
This is a function of the doubled phase-space variables ($q,\wt{p},\wt{q},p)$, and time $t$.  Setting $\wh{R} = \frac{1}{(2\pi\hbar)^2}\wh{\rho}$, we get the Wigner quasi-probability distribution
\begin{align}\label{eq:wignerreprho}
	\rho_W(q,\wt{p},\wt{q},p,t) \coloneqq
	\frac{1}{(2\pi\hbar)^2} W[\wh{\rho}]~.
\end{align}
We immediately see that integrating out $\wt{q}$ and $\wt{p}$ produces the Liouville distribution $\rho_{_L}$
\begin{align}\label{eq:rhoqpmarginal}
	\iint_{-\infty}^{\infty} d\wt{q} \, d\wt{p} \,
	\rho_W(q,\wt{p},\wt{q},p,t) = \rho_{_L}(q,p,t)~.
\end{align}
This is in analogy with quantum mechanics where the marginal of the Wigner quasi-probability distribution produces a true probability distribution.  In the quantum case, the marginal is the modulus-square of the wavefunction, in our case it is the Liouville probability distribution.

We say ``quasi''-probability because $\rho_W$ here has identical properties to that of the quantum Wigner distribution, except that it depends on double the number of variables, and pairs $q$ to $\wt{p}$ and $\wt{q}$ to $p$, instead of $q$ to $p$.  These properties are as follows
\begin{subequations}\label{eq:wignerproperties}
\begin{align}
	1 &=
	\iiiint_{-\infty}^{\infty}
	dq \, d\wt{p} \, d\wt{q} \, dp \, \rho_W~,\\
	\langle \wh{R} \rangle &=
	\iiiint_{-\infty}^{\infty}
	dq \, d\wt{p} \, d\wt{q} \, dp \, \rho_W R_W~,\\
	0 &\leq
	\iiiint_{-\infty}^{\infty}
	dq \, d\wt{p} \, d\wt{q} \, dp \, \rho_W \rho'_W
	\leq \frac{1}{(2\pi\hbar)^2}~,\\
	\frac{1}{(2\pi\hbar)^2} |\langle \chi' | \chi \rangle|^2 &=
	\iiiint_{-\infty}^{\infty}
	dq \, d\wt{p} \, d\wt{q} \, dp \, \rho_W \rho'_W~,
\end{align}
\end{subequations}
where in the last equality, $\rho_W = |\chi\rangle\langle\chi|$ and $\rho'_W = |\chi'\rangle\langle\chi'|$ are pure states.  These properties can be derived directly from the properties~\eqref{eq:stateoperator} and~\eqref{eq:rhoabinequality} of the state operator, and the definition~\eqref{eq:wignerrep} and~\eqref{eq:wignerreprho} of the Wigner representation.  The derivation is identical (other than the doubled number of variables) to its analogue in quantum mechanics~\cite{ballentine_quantum_2014}.  The first of~\eqref{eq:wignerproperties} is the normalization of $\rho_W$, the second shows that it can be used to calculate averages, and the third, if we set $\rho'_W = \rho_W$, that it cannot be too sharply peaked and its area of support cannot be less than $(2\pi\hbar)^2$.  The left-hand-side of the fourth equation vanishes if $|\chi\rangle$ and $|\chi'\rangle$ are orthogonal, which implies that $\rho_W$ cannot be nonnegative in general; this is termed Wigner negativity.

The equation of motion for the Wigner distribution $\rho_W$ can, in principle, be derived from that of the state operator~\eqref{eq:rhoeom}.  In QM, that equation may be stated in terms of a star product (see Refs.~\cite{Hancock2004, Zachos2005, Curtright2014} and references therein).  We will not go through the derivation, but will argue that the only difference between this classical setup and the quantum one is the number (and symbols) of the degrees of freedom ($q$ and $\wt{q}$ for CM, and $q$ for QM) and the form of the generator of time-evolution (the Liouvillian for CM and the quantum Hamiltonian for QM).  Neither of those differences should change the star-product approach, and so we will write the equation of motion for the classical $\rho_W$ in analogy with the quantum one as
\begin{align}
	\partial_t \rho_W + \frac{1}{i\hbar} \left(
	\rho_W \star \wt{H}_W - \wt{H}_W \star \rho_W \right) = 0~,
\end{align}
where the star product is defined as
\begin{align}
	f(q,\wt{p},\wt{q},p,t) \star g(q,\wt{p},\wt{q},p,t) &\coloneqq
	f \, \exp{\left[ \frac{i\hbar}{2} \left(
	\overleftarrow{\partial}_q \overrightarrow{\partial}_{\wt{p}} -
	\overleftarrow{\partial}_{\wt{p}} \overrightarrow{\partial}_q +
	\overleftarrow{\partial}_{\wt{q}} \overrightarrow{\partial}_p -
	\overleftarrow{\partial}_p \overrightarrow{\partial}_{\wt{q}}
	\right) \right]} \, g~.
\end{align}
Here, the left- and right-acting derivatives are defined such that $f \overleftarrow{\partial}_x \overrightarrow{\partial}_y g = \partial_x f \partial_y g$, and the exponential in terms of its series expansion.

Finally, we saw how the Liouville distribution $\rho_{_L}$ was produced as the $qp$-marginal of $\rho_W$ by integrating out $\wt{q}$ and $\wt{p}$.  Another marginal may be of interest; that produced by integrating out $\wt{q}$ and $p$.  We have repeatedly mentioned that $q$ is paired to $\wt{p}$ through the fact that the latter is the generator of translations along the former, implying a canonical commutation relation and an uncertainty relation between them.  We have also argued, from the requirement of the equivalence between KvN and Liouville mechanics, that $q$ and $\wt{p}$ belong to their own Hilbert space $\mathcal{H}_1$, separate from $\mathcal{H}_2$ which contains $\wt{q}$ and $p$.

Define the Wigner representation of operators $\wh{R}^{(1)}$ on $\mathcal{H}_1$ as
\begin{align}
	R^{(1)}_W \coloneqq W^{(1)}[\wh{R}^{(1)}]
	\coloneqq \int_{-\infty}^{\infty} dq' \,
	\langle q - q'/2 | \, \wh{R}^{(1)} \, | q + q'/2 \rangle
	e^{\frac{i}{\hbar} \wt{p} q'}~.
\end{align}
It is easy to show that, if $\wh{\rho}\,^{(1)} = \Tr^{(2)} \wh{\rho}$ is the state operator on $\mathcal{H}_1$, then the $q\wt{p}$-marginal of $\rho_W$ is equal to $\frac{1}{2\pi\hbar}W^{(1)}[\wh{\rho}\,^{(1)}]$
\begin{align}
	\iint_{-\infty}^{\infty} d\wt{q} dp \, \rho_W
	= \frac{1}{2\pi\hbar} W^{(1)}[\wh{\rho}\,^{(1)}] \eqqcolon \rho^{(1)}_W(q,\wt{p},t)~.
\end{align}
Since $\wh{\rho}\,^{(1)}$ is a state operator, its Wigner representation possesses similar properties to those of the full Wigner representation~\eqref{eq:wignerproperties}, including Wigner negativity.  The same setup can be done on $\mathcal{H}_2$ for $\wt{q}$ and $p$.

The partial Wigner representation $\rho^{(1)}_W(q,\wt{p},t)$ then is a quasi-probability distribution on the phase-space constructed from $q$ and its generator of translations $\wt{p}$.  This, judging by its negativity and the character of $q$ and $\wt{p}$, looks more like an analogue of the quantum Wigner function on the quantum phase-space than the Liouville distribution $\rho_{_L}(q,p,t)$ does.\footnote{
	For a different approach, relating the Koopman-von Neumann probability amplitude to the classical limit of the quantum Wigner function, see~\cite{Bondar2012, bondar_wigner_2013}.
}

\section{Conclusion}\label{sec:conc}

Classical mechanics, as shown here, is capable of reproducing two features often regarded characteristic of quantum mechanics.  Importantly, it does so without imposing external rules, such as an epistemic restriction~\cite{spekkens_defense_2007, bartlett_reconstruction_2012, spekkens_quasi-quantization_2016}.  These features, however, appear only when considering tilde-variables, in addition to ordinary phase-space variables.  Just as in quantum mechanics, classical mechanics contains also generators of transformations.  Unlike quantum mechanics, however, classical mechanics considers position and momentum separately, and allows only generators of canonical transformations.  Formalizing the structure of both theories in a list of postulates using the same mathematical formalism allows for a more focused characterization of the similarities and differences between them.

Tethering the Hilbert-space description of classical mechanics to its phase-space one at all instances grounds the classicality of the obtained results.  It shows us that an ordinary classical ontic model could, at least in principle, underlie some features present in quantum mechanics, including uncertainty and Wigner negativity.  The price to pay is to view the operator named ``momentum'' in quantum mechanics as its true definition: the generator of position translations.  This suggests the possibility of comparing quantum momentum not to classical momentum, but to the classical ``tilde-momentum.''  Moreover, it suggests comparing the quantum Wigner function to a marginal of the classical Wigner function.

It is shown here that concepts like uncertainty and Wigner negativity are not, as traditionally thought, foreign to classical mechanics.  Rather, they are unsurprising, once we recognize that classical mechanics have always contained, as dynamical variables, generators of canonical transformations that are fundamentally incompatible with phase-space variables.

\section*{Acknowledgements}
I thank Mark A. Walton for support and guidance throughout the preparation of this manuscript.   This research we supported in part by a Discovery Grant (RGPIN-2022-04225) from the Natural Sciences and Engineering Research Council (NSERC) of Canada.

\AtNextBibliography{\footnotesize}
\printbibliography

\end{document}